

\magnification=1200
\hsize = 5.5 true in
\hoffset = 0.5 true in
\baselineskip=20pt
\lineskiplimit=1pt
\lineskip=2pt plus .5pt
\parskip = 3pt
\font\tenrm=cmr10
\font\ninerm=cmr9

\def\today{\number\month /\number\day /\number\year}
\pageno=0
\headline={\rm \ifnum \pageno<2 LA-UR-94-431 \hfil \today
		\else LA-UR-94431
``Models of Charge Displacements in Water'' \hfil \today \fi}
\footline={\rm \ifnum \pageno<1 \hfil \else \hfil \folio \hfil \fi}

\topskip = 48pt
\parindent =  20pt
\vfil
\centerline{\bf Tests of Dielectric Model Descriptions of Chemical}
\centerline{\bf Charge Displacements in Water}

\bigskip

\centerline{Gregory J. Tawa and Lawrence R. Pratt}
\centerline{Theoretical Division, }
\centerline{Los Alamos National Laboratory, }
\centerline{Los Alamos, NM 87545}

\bigskip
\bigskip


\parindent =  20pt
\topskip = 10pt

\baselineskip=11pt
\ninerm

A dielectric model of electrostatic solvation is applied to describe
potentials of mean force in water along reaction paths for: a)
formation of a sodium chloride ion pair; b) the symmetric SN2 exchange
of chloride in methylchloride; and c) nucleophilic attack of
formaldehyde by hydroxide anion.  For these cases simulation and XRISM
results are available for comparison.  The accuracy of model
predictions varies from spectacular to mediocre.  It is argued that:
a) dielectric models are physical models, even though simplistic and
empirical; b) their successes suggest that second-order perturbation
theory is a physically sound description of free energies of
electrostatic solvation; and c) the most serious deficiency of the
dielectric models lies in the definition of cavity volumes.
Second-order perturbation theory should therefore be used to refine
the dielectric models.  These dielectric models make no attempt to
assess the role of packing effects but for solvation of classical
electrostatic interactions the dielectric models sometimes perform as
well as the more detailed XRISM theory.

\tenrm
\baselineskip=20pt

An important quality of water as a solvent is its ability to stabilize
ions and polar molecules.  Since displacement of electric charge is
often central to chemical reactivity, water is a special solvent for
chemical reactions in solution.  An accurate molecular theory of water
participation in chemical reactions in aqueous solution has not been
established but a few possibilities are available.  The range of
theoretical approaches includes simulation calculations, integral
equation theories, and dielectric models.

	Comparison of the predictions of dielectric models with those of
other methods in cases where thermal precision in solvation free
energis is possible is the goal of this paper.  For several molecular
complexes we obtain thermally accurate solutions of the governing
macroscopic Poisson equation.  The particular examples are chosen
because of the availability of results from alternative methods.  The
comparison of dielectric model results with data from simulation
calculations or with XRISM (1) results should teach us about the
utility of the dielectric model and about fruitful directions for the
discovery of better theories.  Since the goal is unambiguous
comparisons that test the performance of the dielectric model, the
example problems are not discussed for their own sake.

\medskip\noindent {\bf Dielectric models of electrostatic solvation
free energies}

	The dielectric model we apply is physically viewed as follows (2, 3).
Attention is focused on a solute of interest.  A solute volume is
defined on the basis of its geometry.  Partial charges describing the
solute electric charge distribution are positioned with respect to
this volume.  For liquid water under the most common conditions it is
known that the van der Waals volume of the molecule is a satisfactory
choice for the molecular volume (4, 5).  But that is coincidental and
elaborations of the theory will have to consider more general
possibilities.  For the present applications, it is essential that the
defined solute volume permit disconnection when the solute fragments
are widely separated.  We define the solute volume as the volume
enclosed by spheres centered on solute atoms.  The solvent is
idealized as a continuous dielectric material with dielectric constant
$\varepsilon$.  The value $\varepsilon=77.4$, appropriate to water at
its triple point, is used everywhere below.  The solvent is considered
to be excluded from the solute volume and that region is assigned a
dielectric constant of one, $\varepsilon=1$.

\bigskip \noindent {\bf Methods}.  The equation to be solved for the
model is $$\nabla \bullet \varepsilon (r)\nabla \Phi (r)=-\;4\pi \rho
_f(r) \eqno{(1)}$$ where $\rho_f ({\bf r})$ is the density of electric
charge associated with the solute mole\-cule, the function
$\varepsilon({\bf r})$ gives the local value of the dielectric
constant, and the solution $\Phi({\bf r})$ is the electric potential.
To solve this equation, we first cast it as an integral equation, {\it
e. g.\/} $$\Phi (r)=\Phi _0(r)+\int\limits_V {G_0(r,r')\left(
{{{\nabla '\varepsilon (r')} \over {4\pi \varepsilon (r')}}}
\right)\bullet \nabla '\Phi (r')d^3r'}.  \eqno{(2)} $$ Here
$G_0(r,r')$ is the Green function for the Poisson equation with
$\varepsilon (r)=1$ and $\Phi _0(r)$ is the electrostatic potential
for that case.  It is assumed that all the charges of $\rho_f ({\bf
r})$ are positioned in regions where $\varepsilon (r)=1$. This
equation is correct both for a localized distribution $\rho_f ({\bf
r})$ and zero boundary data on a surface everywhere distant and for
periodic boundary conditions on a cell of volume $V$.  $G_0(r,r')$ is
different in those two cases as is $\Phi _0(r)$.  This equation is not
the only such form that can be solved and more general considerations
can be helpful.  But we do not pursue those issues here.

	The integrand of equation 2 is concentrated on the interface between
the solute volume and the solvent.  We can then use boundary element
ideas to solve it (6-11).  The principal novelty in our numerical
methods is that we use a sampling method based upon quasi-random
number series (12) to evaluate the surface integral rather than more
specialized methods.  Advantages of our method are that it facilitates
systematic studies of numerical convergence and exploitation of
systematic coarse-graining.  More specific discussion of numerical
methods can be expected at a later date.

	With the solution of equation 2 in hand we obtain the desired
potential of mean force as $$W=U+\left( {{1 \over 2}}
\right)\int\limits_V {\rho _f(r)\left( {\Phi (r)-\Phi _0(r)}
\right)d^3r} \eqno{(3)} $$ where $U$ is the static energy in the
absence of the solvent.  Since for the present examples $\rho_f ({\bf
r})$ is a sum of partial charges, the integral in equation 3 is a sum
over those partial charges.

\noindent {\bf Results}

\noindent {\bf a) Pairing of sodium and chloride ions in water}.
Dielectric model results for the $Na^+ \cdots Cl^-$ potential of
average force in water are shown in Figure 1.  The radii used were
those recommended by Rashin and Honig (5).  See also Pratt, {\it et
al.\/} (Pratt, L. R., Hummer, G., and Garcia, A. E., {\it Biophys.
Chem.\/}, in press).  These results agree with those of Rashin who
studied a similar dielectric model (13).  Rashin assumed a somewhat
different solute volume and his predicted potential of mean force
displayed a more prominent barrier to escape from the contact minimum.
Those results were similar to XRISM (14).  The dielectric model
results are surprising in showing minimum free energies both at ion
contact and at a larger distance that indicates a solvent-separated
pairing.  Although surprising, these results are not in quantitative
agreement with simulation calculations of solvation free energies of
ion pairing (Hummer, G., Soumpasis, D. M., and Neumann, M., {\it Mol.
Phys.\/}, in press).  Most importantly, the contact minimum is much
deeper than the simulation results.  We note also that the simulation
results were obtained for finite concentrations of NaCl.  Thus, the
large distance behavior of that potential of mean force is influenced
by ionic screening.  The dielectric models and this XRISM result
conforms to the asymptotic variation expected at infinite dilution.

A particular concern over recent years has been sensitivity of the
molecular results for potentials of mean force to modelled
intermolecular interactions (15).  Figure 1 does not attempt to give
the wide range of results that have been obtained.  However, to the
extent that such sensitivity is considered important, it highlights a
fundamental deficiency of the dielectric models.  Those molecular
details are not present in the dielectric models as currently applied.
The dielectric models do depend on the dielectric constant of the
solvent but for many applications that dependence is not decisive
because of the high values of the dielectric constant that are
typically relevant.

\noindent {\bf b) Symmetric S$_N$2 chloride exchange in methyl
chloride}.  Following the efforts of Chandrasekhar, {\it et al.\/}
(16), this example has been taken as a theoretical model of reactions
in solutions over recent years (17-18).  We used radii of $R_H =
1.00$\AA, $R_C = 1.85$\AA, $R_{Cl} = 1.937$\AA, and partial charges of
Chandrasekhar, {\it et al.\/} (16) (Figure 2) and Huston, et al. (18)
(Figure 2 inset).  As Figure 2 shows, the agreement between dielectric
model and simulation results is close, no less satisfactory than the
agreement between XRISM and simulation.  As is discussed in the paper
of Huston, {\it et al.\/} (18) the original assignment of partial
charges of the reacting complex in the neighborhood of the barrier
leads to the prediction of the notch in the potential of the mean
force by the XRISM theory.  This is also true of the dielectric model,
as is seen in Figure 2.  Huston, {\it et al.\/} (18) reanalyzed this
modeling of solute-solvent interactions and proposed alterations of
the original parameterization.  The results for this alternative
description of the barrier region are shown in the inset.

\noindent {\bf c) Nucleophilic attack of $HO^-$ on $H_2CO$}.  This
example has been previously studied as a prototype of an initial step
in the formation and destruction of the peptide unit (19, 20).  We
used radii of $R_{O(hydroxide)} = 1.65$\AA, $R_C = 1.85$\AA, and $R_H
= 1.0$\AA, and $R_{O(carbonyl)} = 1.60$\AA, and the partial charges
and geometries of Madura and Jorgensen (19).  The results are shown in
Figure 3.  The predictions of the dielectric model are qualitatively
similar to those of the Monte Carlo simulation and of the XRISM
theory, particularly in predicting the existence of a solvation
barrier prior to contact of the reacting species.

The results of the dielectric model are not quantitatively accurate in
this case, however.  It seems clear that adjustment of the cavity
radii at each geometry could bring the model results into agreement
with the simulation data.  Recently, a new proposal was given for
achieving a physically valid parameterization of the model on the
basis of additional molecular information (Pratt, L. R., Hummer, G.,
and Garcia, A. E., {\it Biophys.  Chem.\/}, in press); that proposal
is discussed below.  As an observation preliminary to attempts to
parameterize the dielectric model on more basic information, we
demonstrate here that reasonable radii can achieve excellent agreement
with the data.  To do this we adjusted the radii of the hydroxide
oxygen to reproduce at a coarse level of calculation the simulation
results at a few points along the potential of mean force shown, used
a simple interpolation for radii at other points, and observed the
variation of the hydroxide oxygen radius.  The adjusted radii are
shown in Figure 4 and the results for the potential of mean force are
the circles of Figure 3.  This type of comparison can be misleading.
Agreement just as good would be achieved even if the model result were
physically wrong, {\it e. g.\/}, if we had mistakenly done the
calculation for acetaldehyde rather than formaldehyde.  However, the
typical rough expectation for cavity radii for these models is between
the van der Waals radius and the distance of closest approach for
solvent molecular centers.  To the extent that the radii are treated
as truly adjustable, the model has more than enough flexibility to
match the data with reasonable changes in those radii.

\bigskip\noindent {\bf Discussion}

\noindent The importance of the dielectric models is that they provide
free energies of solvation for cases that are not accessible by
alternative methods.  The free energies obtained from the dielectric
model are quadratic functionals of the charge distribution of the
solute.  It has been pointed out previously (Pratt, L. R., Hummer, G.,
and Garcia, A. E., {\it Biophys.  Chem.\/}, in press).  that this
indicates that the dielectric models correspond to a modelistic
implementation of second-order perturbation theory for the excess
chemical potential of the solute.  Thus, the successes of dielectric
models suggest that second-order thermodynamic perturbation theory is
a physically sound theory for the desired solvation free energy due to
electrostatic interactions.  But in addition to the Tsecond-orderU
limitation, dielectric models also drastically eliminate molecular
detail of the solvation structures.  This detail can be restored by
implementing second-order thermodynamic perturbation theory on a
molecular basis.  The fundamental formula for that approach is
$$\Delta \mu \approx \Delta \mu _0+\left\langle
{\sum\limits_{C=constituents} {\varphi (C)}} \right\rangle _0 $$ $$
-\left( {{\beta \over 2}} \right)\left\langle {\left(
{\sum\limits_{C=constituents} {\varphi (C)}-\left\langle
{\sum\limits_{C'=constituents} {\varphi (C')}} \right\rangle _0}
\right)^2} \right\rangle _0. \eqno{(4)}$$ This approximation has been
discussed previously by Levy, {\it et al.\/} (23).  Here the subscript
`0' indicates quantities obtained for the reference system in which no
electrostatic interactions are expressed between the solution
constituents and a designated solute molecule.  The $\varphi (C)$ are
the electrostatic potential energies of interaction between the
constituent $C$ of the solution and that solute.  Several aspects of
this molecular approach should be noted.  First, it requires knowledge
of $\Delta \mu _0$, the solvation free energy when electrostatic
interactions are neglected.  This is not supplied by the dielectric
models.  Second, the molecular approach includes a term linear in the
charges; this involves the potential at zero charge induced by short
ranged forces.  This term is generally present, non-zero, dependent
upon molecular geometry and thermodynamic state, but the dielectric
models assume that it vanishes.  Third, the second-order term
corresponds to the quantities usually obtained from dielectric models
but this formula avoids the classic empirical adjustments of cavity
radii.  From the perspective of a continuum approach, this second
order term incorporates nonlocality of the polarization response of
the solution.

	In view of the corresponding molecular theory, a natural way to
improve dielectric model results is first to obtain molecular results
for the second-order term to better establish the solute volume on a
proper molecular basis.  Cavity radii thus determined will be
dependent upon the thermodynamic state just as empirical cavity radii
must generally be considered functions of thermodynamic state.  For
example, evaluation of enthalpies by temperature differentiation
should include derivatives of the cavity radii with respect to
temperature (22).  After definition of that solute volume is better
controlled, an alternative source of information on the leading two
terms must be developed.  At this level, the cavity radii are
independent of charges and charge distributions of the solute.
Finally, the importance of succeeding terms in the perturbation theory
must be assessed (Pratt, L. R., Hummer, G., and Garcia, A. E., {\it
Biophys.  Chem.\/}, in press, Rick, S. W., and Berne, B. J., {\it J.
Am. Chem. Soc.\/}, in press ).  Those succeeding terms are likely to
be especially troublesome for circumstances where composition
fluctuations are physically important, {\it e. g.\/}, for mixed
solvents.

	We note again that the reactions and models studied here were chosen
solely because of the availability of molecularly detailed simulation
data and of results of integral equation theory.  Since testing of the
dielectric model by comparing its predictions to simulation and
integral equation results is the objective, the same model systems
must be treated by the various methods.  Solute geometries and partial
charges must be accepted for the test.  Because of the simplicity of
the dielectric model, however, we do not expect that conclusions
regarding the utility of the model for description of charge
displacement in aqueous solutions would be significantly changed if
more elaborate models of these reactions were available.  For example,
the generalization of equation 4 to apply when solute-solvent
electrostatic interactions are described with the aid of higher order
multipole moments of the solute charge distribution is
straightforward.

\noindent {\bf Conclusions}

	To the extent that the dielectric model is physically sound,
second-order thermodynamic perturbation theory should provide an
accurate description of free enrgies due to electrostatic interactions
between the solute and the solution.  Second-order thermodynamic
perturbation theory restores molecular detail of the solvation
structures that is discarded when the dielectric model is used.  More
fundamentally, second-order thermodynamic perturbation theory
identifies the potential at zero charge that is neglected in the
dielectric model.

	Despite their simplistic character, dielectric models provide a
physically sound description of chemical charge displacements in
water.  Because of these qualities they can be helpful where only
rough but physical results are required; for example, they might be
expected to provide serviceable umbrella functions (23) for more
accurate molecular calculations of free energies along reaction paths
of the sort considered here.  It should be recognized, however, that
stratification of the reaction coordinate is typically more important
(24, 25).

	Considered directly the dielectric models are not reliably accurate
for thermal level energy changes.  A large part of the unreliability
of the dielectric model predictions is surely due to the assignment of
cavity radii; the predictions of the model are sensitive to those
parameters, they clearly ought to vary along a reaction path, and the
determination of proper values for the radii comes from outside the
model.  Although the dielectric model does not attempt to assess the
importance of packing effects on solvation properties, it is sometimes
of comparable accuracy to the more detailed XRISM theory for treatment
of electrostatic contributions.  The dielectric model also has the
advantage of being simple and physical when used for those purposes.

\noindent {\bf Acknowledgments}.  We are grateful for helpful
discussions with Drs. J. Blair, S.-H. Chou, P.  Leung, D. Misemer, J.
Stevens, and K. Zaklika of 3M Corporation on the topics of solvation
and reaction chemistry in solution.  LRP thanks Gerhard Hummer and
Angel E. Garcia for helpful discussions and acknowledges partial
support for this work from the Tank Waste Remediation System (TWRS)
Technology Application program, under the sponsorship of the U. S.
Department of Energy EM-36, Hanford Program Office, and the Air Force
Civil Engineering Support Agency.  This work was also supported in
part by the US-DOE under LANL Laboratory Directed Research and
Development funds.

\noindent {\bf Literature Cited}

\item{ 1.} Rossky, P. J., {\it Ann. Rev. Phys. Chem.\/} {\bf 1985} ,
36, 321.

\item{ 2.} Rashin, A. A., {\it J. Phys. Chem.\/} {\bf 1990}, 1725.

\item{ 3.} Honig, B., Sharp, K., and Yang, A.-S., {\it J. Phys.
Chem.\/} {\bf 1993}, 97, 1101.

\item{ 4.} Latimer, W. M., Pitzer, K. S., and Slansky, C. M., {\it J.
Chem. Phys. \/} {\bf 1939}, 7,108.

\item{ 5.} Rashin, A. A., and Honig, B., {\it J. Phys. Chem.\/} {\bf
1985}, 89, 5588.

\item{ 6.} Pascual-Ahuir, J. L., Silla, E., Tomasi, J., Bonaccorsi,
{\it J. Comp. Chem.\/} {\bf 1987}, 8, 778.

\item{ 7.} Zauhar, R. J., and Morgan, R. S., {\it J. Mol. Biol.\/}
{\bf 1985}, 186, 815; {\it J. Comp.  Chem.\/} {\bf 1988}, 9, 171.

\item{ 8.} Rashin, A. A., and Namboodiri, K., {\it J. Phys. Chem.\/}
{\bf 1987}, 91, 6003.

\item{ 9.} Yoon, B. J., and Lenhoff, {\it J. Comp. Chem.\/} {\bf
1990}, 11, 1080; {\it J. Phys. Chem.\/} {\bf 1992}, 96, 3130.

\item{ 10.} Juffer, A. H., Botta, E. F. F., van Keulen, A. M., van der
Ploeg, A., and Berendsen, H. J. C., {\it J. Comp. Phys.\/} {\bf 1991},
97, 144.

\item{ 11.} Wang, B., and Ford, G. P., {\it J. Chem. Phys.\/} {\bf
1992}, 97, 4162.

\item{ 12.} Hammersley, J. M., and Handscomb, D. C., {\it Monte Carlo
Methods\/}; Chapman and Hall: London, 1964; pp. 31-36.

\item{13.} Rashin, A. A., {\it J. Phys. Chem.\/} {\bf 1989}, 93, 4664.

\item{ 14.} Pettitt, B. M., and Rossky, P. J., {\it J. Chem. Phys.\/}
{\bf 1986}, 84, 5836.

\item{ 15.} Dang, L. X., Pettitt, B. M., and Rossky, P. J., {\it J.
Chem. Phys.\/} {\bf 1992}, 96, 4046.

\item{ 16.} Chandrasekhar, J., Smith, S. F., and Jorgensen, W. L.,
{\it J. Am. Chem. Soc.\/} {\bf 1985}, 107, 154.

\item{ 17.} Chiles, R. A., and Rossky, P. J., {\it J. Am. Chem.
Soc.\/} {\bf 1984}, 106, 6867.

\item{ 18.} Huston, S. E., Rossky, P. J., and Zichi, D. A., {\it J.
Am. Chem. Soc.\/} {\bf 1989}, 111, 5680.

\item{ 19.} Madura, J. D., and Jorgensen, W. L., {\it J. Am. Chem.
Soc.\/} {\bf 1986}, 108, 2517.

\item{ 20.} Yu, H.-A., and Karplus, M., {\it J. Am. Chem. Soc.\/} {\bf
1990}, 112, 5706.

\item{ 21.} Levy, R. M., Belhadj, M., and Kitchen, D. B., {\it J.
Chem. Phys.\/} {\bf 1991}, 95, 3627.

\item{ 22.} Roux, B., Yu, H.-A., and Karplus, M., {\it J. Phys.
Chem.\/} {\bf 1990}, 94, 4683.

\item{ 23.} Valleau, J. P, and Torrie, G. M., in {\it Statistical
Mechanics, Part A: Equilibrium Techniques\/}; Berne, B. J., Ed.;
Modern Theoretical Chemistry; Plenum: New York, New York, 1977, Vol.
5; pp 178-182.

\item{ 24.} Hammersley, J. M., and Handscomb, D. C., {\it Monte Carlo
Methods\/}; Chapman and Hall: London, 1964; pp 55-57.

\item{ 25.} Kalos, M. H., and Whitlock, P. A., {\it Monte Carlo
Methods Volume I: Basics\/}; Wiley-Interscience: New York, 1986; pp
112-115.

\bigskip\noindent {\bf Figure Captions}

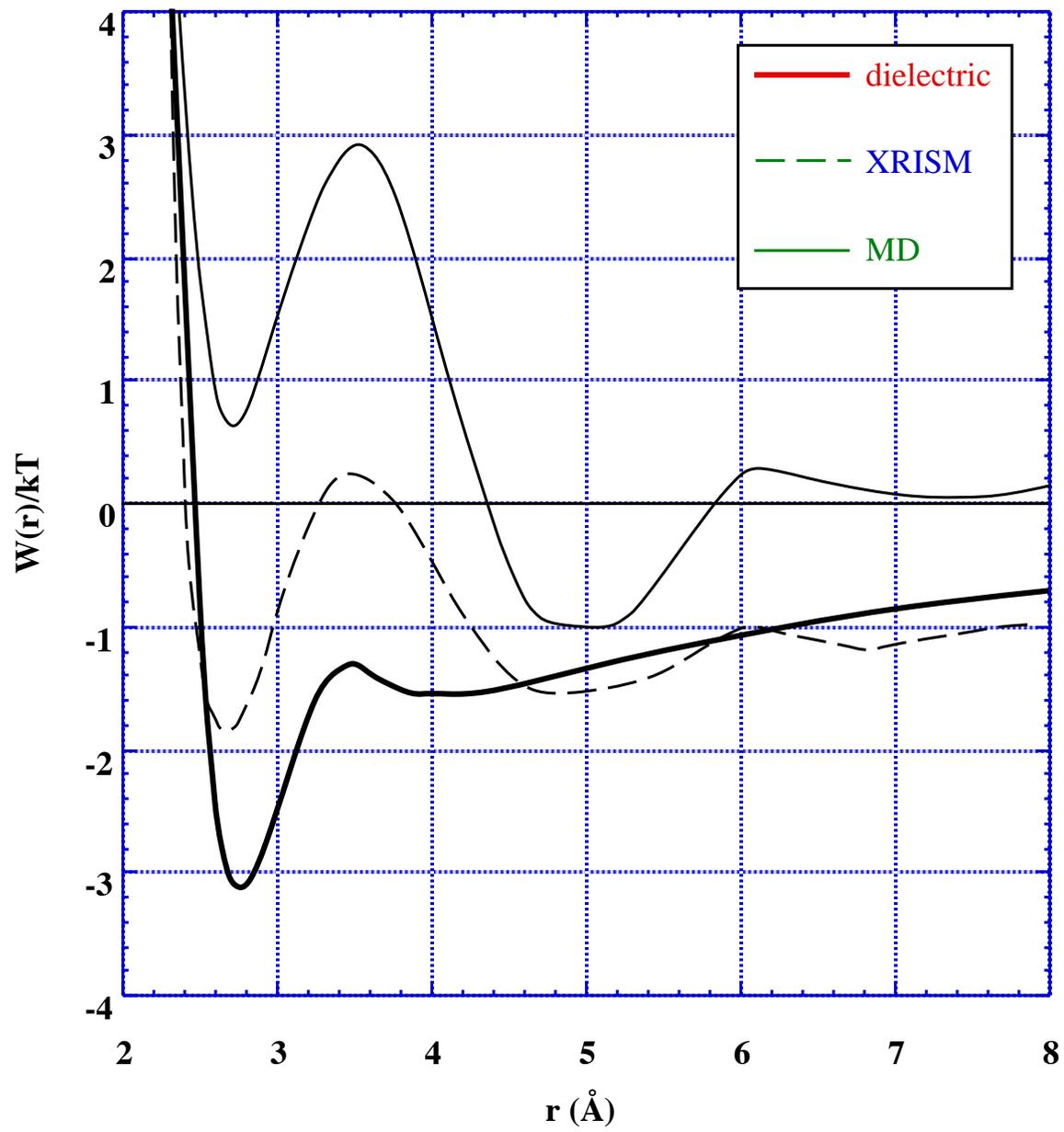
\item{Figure 1:}Potentials of the mean forces between ion pairs Na$^+
\cdots$Cl$^-$ in water.  The XRISM results are redrawn from Reference
14; the MD results are redrawn from Pratt, {\it et al.\/} (Pratt, L.
R., Hummer, G., and Garcia, A. E., {\it Biophys. Chem.\/}, in press).
Those original MD results are to be found in Hummer, {\it et al.\/}
(Hummer, G., Soumpasis, D. M., and Neumann, M., {\it Mol. Phys.\/}, in
press).

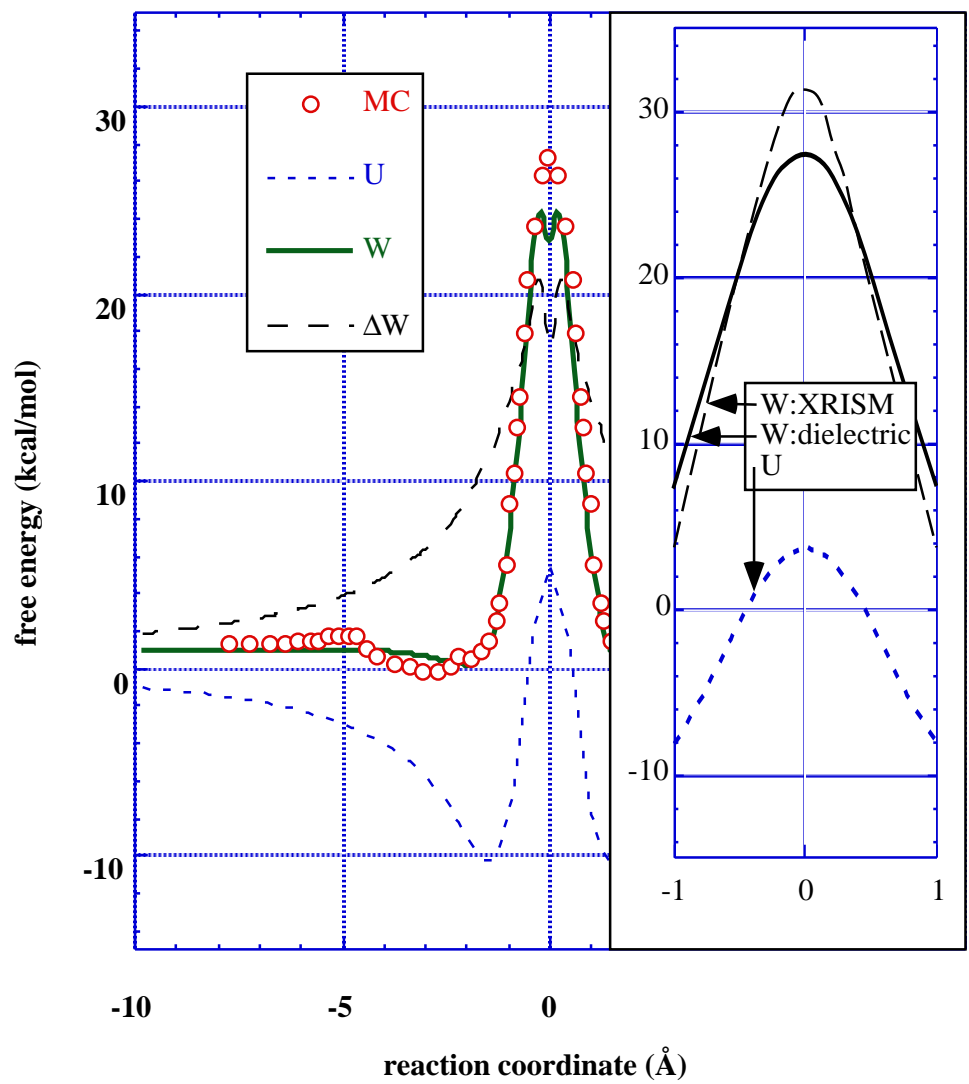
\item{Figure 2:} Potential of the mean force along a reaction path for
symmetric $S_N2$ replacement of the Cl$^-$ ion in CH$_3$Cl.  $\Delta W
\equiv W - U$.  See References 16 and 18.

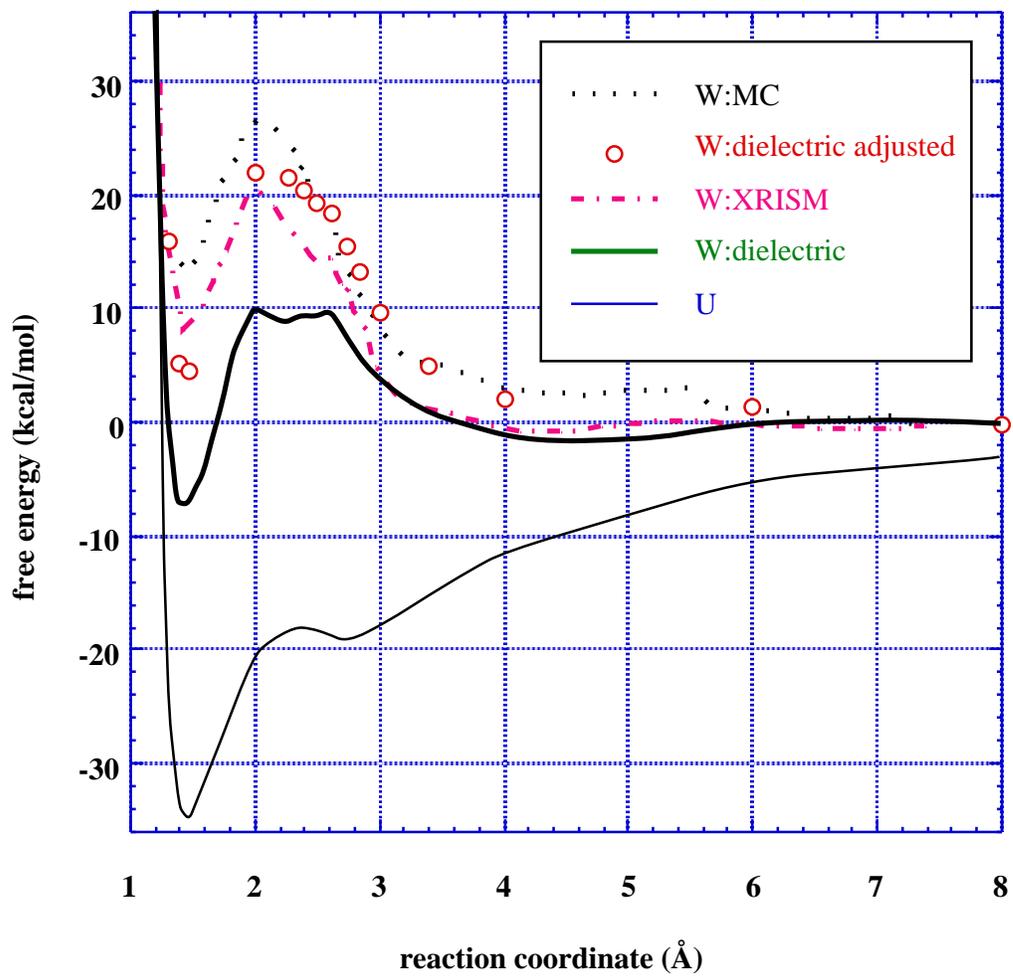
\item{Figure 3:} Potential of the average force along a reaction path
for nucleophilic attack of H$_2$CO by HO$^-$ according to the
dielectric model, simulation (19), and XRISM theory (20).  The circles
are the results of the dielectric model with empirical adjustment of
the radius of the hydroxide oxygen.  See the text and Figure 4.

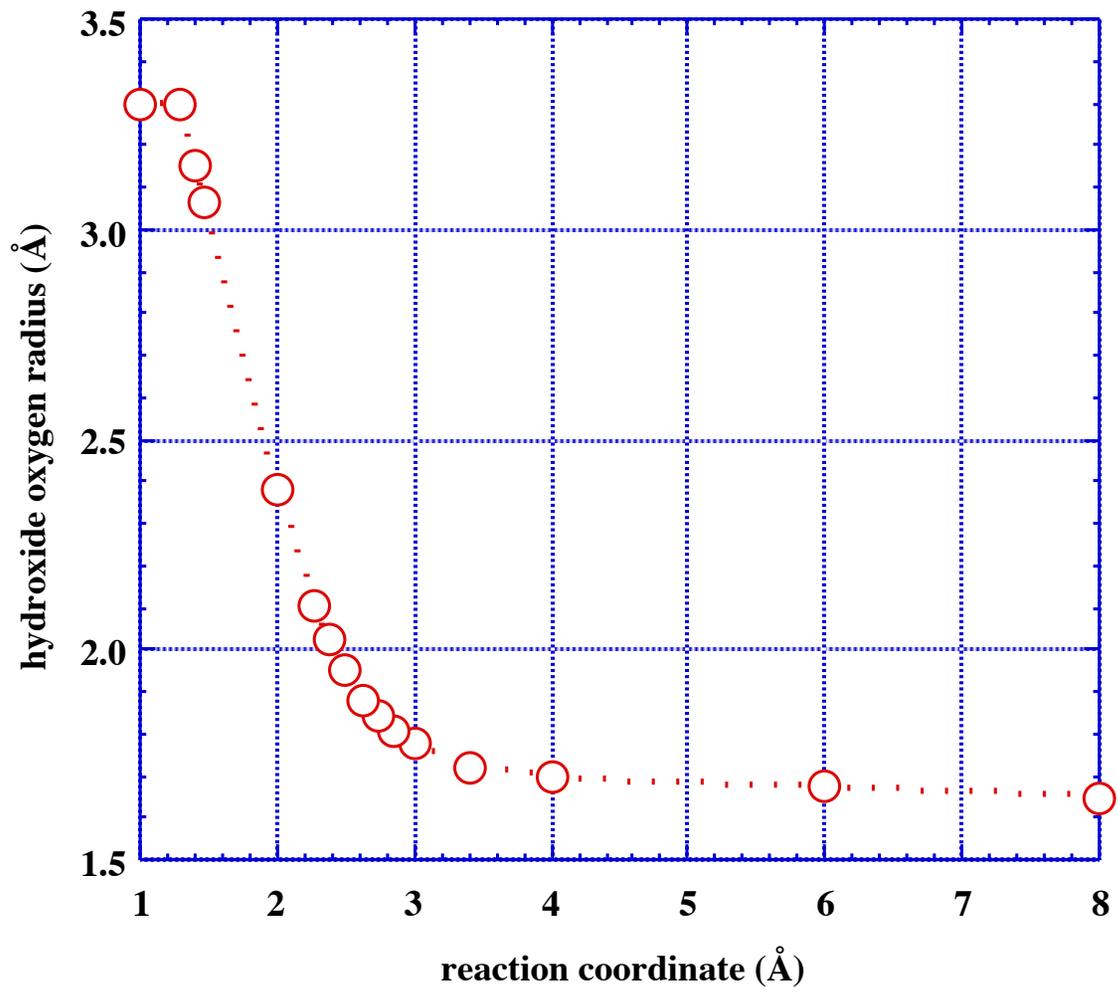
\item{Figure 4:} Variation of the hydroxide-oxygen radius adjusted to
fit the simulation data for the potential of mean force in example
{\bf c}.  See
Figure 3.

\end